\documentclass{article}
\usepackage{hyperref}
\usepackage{amscd,amsmath,amsthm,amssymb}
\usepackage{enumerate,varioref}
\usepackage{epsfig}
\usepackage{graphicx}
\usepackage{mathtools}
\usepackage{tikz}
\newtheorem{thm}{Theorem}

\theoremstyle{definition}
\newtheorem{defn}[thm]{Definition}
\theoremstyle{remark}

\newtheorem{rem}[thm]{Remark}
\numberwithin{equation}{section}
\DeclareMathOperator\arcosh{arcosh}
\newcommand{\R}{\mathbb{R}}

\newcommand{\half}{\frac{1}{2}}

\title{A Family of Metrics for Clustering Algorithms}
\author{Clark Alexander \\Sofya Akhmametyeva \\
Nousot Applied Research Group \\ email \href{mailto:clark@nousot.com}{clark@nousot.com}}

\begin{document}
\maketitle

\begin{abstract}
We give the motivation for scoring clustering algorithms and a metric $M : A \rightarrow \mathbb{N}$ 
from the set of clustering algorithms to the natural numbers which we realize as
\begin{equation}
M(A) = \sum_i \alpha_i |f_i - \beta_i|^{w_i}
\end{equation}
where $\alpha_i,\beta_i,w_i$ are parameters used for scoring the feature $f_i$, which is computed empirically..  We give a method by which one can score features such as stability, noise sensitivity, etc and derive the necessary parameters.  We conclude by giving a sample set of scores. 
\end{abstract}

\section{Introduction}
Data science has been around in form and function if not necessarily in name for the better half of the last century.  Among the main problems that we encounter is ``clustering" which one can describe in layman's terms as grouping elements from a (data)set into subsets which share some feature.  Early uses for data clustering came from the world of particle physics wherein particle accelerators sent large numbers of elementary particles in opposite directions in hopes of some of them colliding with each other to produce nuclear debris.  Several questions arose naturally from this process.  First; how does one read off the data?  Second; if two particles collide, how do we detect this?  Third; is a collision head-on or  are the particles glancing off each other?  Fourth; which data are simply noise? An additional source of data clustering, which arrived later came in the form of astronomy.  One can ask, given a telescope image, how do we discern which spots are stars, galaxies, nebulae, planets, etc?  

Since digital computers were in the nascence in the late 1970s the algorithms needed to operate quickly and stably, with as little sensitivity to noise as possible.  The big constraints then were speed and processing power.  It was difficult to run through a small dataset thousands of times.  This led to algorithms like CLASSY and k-means (reference to 1979 CLASSY article)

In the early decades of the twenty first century running through datasets of as much as a billion points requires very little runtime, and thus thousands of tests is a feasible computational task.  This, however, is buoyed by the fact that datasets are becoming larger and more noisy, to the point where stability and noise sensitivity become even more important.  At present there are dozens, if not hundreds, of well-known algorithms for clustering built for tackling specific problems.  The question naturally arises as to which algorithm is the best.  Perhaps a hybridized approach works for some industries better than others.  Thus we have turned the question of clustering into a question of metrology.  How does one measure the effectiveness of a particular algorithm?  We attempt to answer this question by giving a metric inspired by the points systems for multi-sport events such as pentathlon, heptathlon, and the decathlon.  To measure the athletic performance in such an event, we take into account that there are essentially two types of sports contained within (i) those in which the highest score is the winner, (i.e. pole vault, long jump, javelin), (ii) those in which the lowest score is the winner (i.e. hurdles, 100m, 800m, etc). In sports where the highest score is the winner one wishes to add scores: thus one considers a point value given by
\[
\alpha(score - \beta)^w
\]
In this case, $\beta$ is a reference point, perhaps the lowest qualifying score.  For example in the pole vault, one only considers vaults that get above the 4m mark.  In order to scale the unit into ``points" one uses a scaling factor $\alpha$.  The scaling factor also allows one to scale in such a way that vastly scattered scores (such as long jump measured in cm) does not drastically affect the score versus the 100m sprint measured in seconds.  Finally, the weight $w$ allows one to scale the score so that roughly every event can have roughly the same mean or same variance as whichever criterion an athletic committee would deem more desirable.  One can add more parameters to give the scores some other sort of unity such as maximum range of scores, difference from theoretical maximum, etc.  In our scoring we consider only three parameters, for reference, units, and weighting.

\section{Stability}

Our algorithmic scores are based on their utility for our purposes.  The parameters we derive for our algorithm may be modified for any researcher to meet her/his needs.  For our purposes we wish to have an algorithm with a high amount of stability.  In order to quantify stability, we offer the heuristic definition that given a sufficiently large dataset with $N$ features we shall test some subset $m<N$ features at a time and then exchange one feature at a time.  For example, given a test dataset with 100 features, we shall test approximately 20 at a time.  After we test 20 features we mark the clusters to which each datum belongs.  We then repeat the algorithm while testing 20 features, 19 of which are similar.  We calculate how many points move from their clusters (considering the 20th cluster to be the ``same").  If nearly half of all points move, then we consider this algorithm to score badly in stability.  In fact, we will mark this as zero-stable.  We test this algorithm multiple times and each time count the proportion of points which move from test run to test run.  The average of moving points will be our instability, and thus we define stability of an algorithm $A$ as follows:
\begin{equation}
\text{Stability}(A) = 1 - \frac{\# \text{Moving Points}}{\# \text{Points}} 
\end{equation}

Where the number of moving points is calculated empirically.  Obviously, unless our dataset is sufficiently small, we cannot get the true measure of stability, but we give some confidence interval around the measured mean.  From first year statistics we simply apply the student's $t$-score to find a confidence interval around the measured mean (stability).
\[
\mu \in \left[\hat{\mu}- t\frac{\hat{s}}{\sqrt{n}},\hat{\mu}+ t\frac{\hat{s}}{\sqrt{n}}\right]
\] 

where $n$ is the number of measurements and $\hat{s},\hat{\mu}$ are the measured standard deviation and mean respectively.  Thus we set a threshold of allowable error and calculate the number of experiments necessary to achieve the desired precision.  For our purposes, all datasets will be significantly larger than 30 points which a first course in elementary statistics tells us is a sufficient quantity to estimate with $z$-score.  For the sake of simplicity we shall take $z=2$ and score a 95\% (approximately) confidence interval around our measured stability.

Once we have a quantity called stability, we now scale it to give a maximum of 100 points.  The choice of 100 is completely arbitrary and may be moved to any number to fit a particular researcher's needs.  Our reference point, as mentioned above will be $\half$.  That is to say, if on average half of the points move at every exchanged feature then the algorithm will not be considered stable.  For example, k-means/medians will be considered highly unstable because the choice of initial means/medians dramatical affects the final outcome.  This is particularly noticeable in k-means when considered $m$ dimensional subspaces of $N$ dimensional data.  

\section{Noise}

Noise sensitivity is among the more difficult qualities to quantify with any consistency.  Part of the problem is that we have different types of noise and of utmost importance is the arbitrary nature of the ``shape" of a general dataset.  Thus measuring noise sensitivity requires a little bit of bootstrapping and we will reverse engineer our datasets to test an algorithm's efficacy.  The initial thought in characterizing an algorithm's sensitivity to noise saw the assignment of the score
\[
P(x \text{ causes a problem }| x  \text{ is noise})
\] 

Defining ``causes a problem" is already subjective enough in the sense that this can mean widely varying things for different researchers.  For example, DBSCAN generally handles noise well unless $\varepsilon$ is too large.  In this case, a noisy point can tie together two distinct clusters.  This particular type of error can show up in drawing political district maps.  The second problem with the initial assignment of score is that, if we know a priori that $x$ is, in fact, noise, then we can filter it in preprocessing.  In essence the definition is still too heuristic to be computationally useful; that is, it's not constructive.   

The solution we propose to measuring is noise is to begin with a properly clustered dataset of arbitrary size.  For example we build a spatial dataset which has ``well-defined" clusters of different sizes, shapes, and densities.  Once we have a well-clustered dataset we add noise to it and then recluster.  The main measure then shall be entropy.  More precisely, we will define two probability distributions and measure their Kullback-Leibler divergence.  

First, let's give some definitions and conventions about how we shall calculate sensitivity to noise.  

\begin{defn}
Let 
\[
X = \{x_1,\dots,x_N\}
\]
be a dataset of $N$ points.  Furthermore, let 
\[
Z = \{n_1,\dots,n_q\}
\]
be a set of $q$ points which we consider to be noise.  
Let $X$ be well-clustered with clusters $C_i$.
\[
X = \sqcup_{i} C_i
\]

We shall assign to $X$ the probability mass function which assigns to each cluster its proportional weight.
\[
p_i = |C_i|/N
\]
\end{defn}

The entropy $H$ associated to $X$ is now defined as
\begin{equation}
H(X) = -\sum_i p_i \log(p_i)
\end{equation}

Notice that $p_i<1$ so the negative sign gives us a positive value.  This gives us a good measure of the already existing noise or uncertainty present within $X$.  Now we introduce the Kullback-Leibler divergence
\begin{defn}
Given two probability distributions $p(x)$ and $q(x)$ the Kullback-Leibler divergence is
\begin{equation}
D(p \| q ) = - \sum p(x) \log\left(\frac{p(x)}{q(x)}\right)
\end{equation}
\end{defn}

This gives us a canonical way to measure the ``difference" between two random variables with distributions $p$ and $q$.  In our case, the distribution $p$ will always be the distribution defined by our well-clustered set $X$.

Now we define the sets
\[
Z_{[k]} \subseteq Z
\]
so that $|Z_{[k]}| = k$ has $k$ noisy points.  There are $\binom{q}{k}$ such sets $Z_{[k]}$.  for given $k$. and if we consider all $ k\in \{0,\dots,q\}$
then there are $2^q$ total noisy sets.  Now we create the new sets:
\[
X \cup Z_{[k]}
\]

and we make the probability associations
\[
X \rightsquigarrow p, X \cup Z_{[k]} \rightsquigarrow \tilde{p}_{[k]} 
\]

From here we make the formal definition
\begin{equation}
D(p\|\tilde{p}) = \frac{1}{2^q}\sum_{\text{all } Z_{[k]}} D(p\| \tilde{p}_{[k]}) 
\end{equation}

After having made this formal definition, we realize that it is impossible to calculate this divergence exactly, since we cannot test every possible noisy dataset in existence.  Thus we pick some amount of noisy points and compute several different divergences and again estimate the true divergence against our measured divergence using the $t$-statistic.  

\begin{equation}
\hat{D}(p\|\tilde{p}) \in \left[D(p\|\tilde{p}) - t \frac{\hat{s}_D}{\sqrt{n}},D(p\|\tilde{p}) + t \frac{\hat{s}_D}{\sqrt{n}} \right]
\end{equation}

Thus the larger number of samples $n$ we take, the more accurate our measured divergence is against the true value.  As a quick counting exercise, let's pick 
\[
q = \text{clusters} \times \text{dimensions} := C_X \times d
\]

then we pick $k \approx q/2$.  This gives us a good bound on the number of possible noisy datasets we can choose of size $k$.  

\begin{eqnarray}
|Z_{[k]}| &\approx& \binom{2k}{k} \\
& \approx & \frac{4^k}{\sqrt{\pi k}} \nonumber
\end{eqnarray}

where we have used Stirling's approximation
\[
k! \approx \sqrt{2\pi k}\frac{k^k}{e^k}
\]
to compute the second line. Thus we have approximately
\[
\frac{4^{d C_X}}{\sqrt{\pi d C_X}}
\]

possible noisy datasets containing roughly half the noisy points to test.  Even in low dimensions (i.e. $< 100$) and a small number of clusters, this allows us sufficiently many tests to get a measured divergence to within any chosen accuracy at the cost of running more experiments.

Given that we have a precise measure for divergence we read the consequences of the definition.  Having no divergence means that an algorithm is not susceptible to noise.  The larger the divergence, the higher the susceptibility to noise.  Thus we seek a score in the form
\[
\alpha(\beta - \hat{D}(p\|\tilde{p}))^w
\]

By our choice of $\tilde{p}$ we can get a bound on maximal divergence.  If our algorithm is highly susceptible to noise then each noisy point will create a new cluster.  Then we see
\[
D(p\|\tilde{p}) = -\sum p \ln\left(\frac{p}{\tilde{p}}\right) = -\sum p \ln\left(\frac{p}{k p}\right) = \ln(k) \sum p = \ln(k)
\]

Recalling that we have chosen $k = \frac{d C_X}{2}$ we find a maximal divergence of

\[
\max{D}  = \ln(dC_X) - \ln(2)
\]

In this case we will pick $\beta = \ln(dC_X)$ as our reference point.  This will give $(\beta - D)$ a minimum score of $\ln(2)$ and a maximal score of $\ln(dC_X)$.  We wish to have the raw score as 100 before weighting so we pick $\alpha = 100/\ln(dC_X)$.  

We view noise (in)sensitivity as among the most important features in our metric, thus we give the initial weight $w = 1.25$.  Notice here, that $ln(dC_X)>1$ and thus we wish to increase its overall importance with a weighting factor, thus we pick $w>1.$  It is important to note, however, that divergence is highly sensitive to the initial dataset $X$ and so each researcher must adjust $\alpha,\beta,w$ according to her or his dataset.

\section{Complexity}

Complexity is a well established concept in the theory of computing.  In this case, we shall prefer to use the standard definition of big-Oh.  Our choice of big-Oh rather than big-Theta is based on the fact that we want to score our algorithms based on the slowest that they might be.  Indeed, a different researcher may prefer to use big-Theta, or perhaps for a widely distributed system, clock time. In all cases, the scoring procedure follows in a nearly identical fashion; the faster the better.  In terms of big-Oh we will look only at the exponent.  

\begin{defn}
Consider two functions $f,g:\mathbb{N}\rightarrow \R^+$.  We say 
\[
f(x) \in O(g(x))
\]
if there exists a pair of positive numbers $k,N>0$ so that
\[
\forall x>N, |f(x)| \le k\cdot|g(x)|
\]
\end{defn}  

Notice our slight deviation from the standard $f=O(g)$ to $f\in O(g)$.  This more properly reflects the fact that $O(g)$ is not an equivalence relation since it is not symmetric.  This also allows us to avoid ``tight" bounds.  In essence we are really scoring the highest exponent of $f$ versus the highest exponent of $g$.  A slight reformulation gives us
\begin{equation}
f(x)\in O(g(x)) \iff \lim_{x\rightarrow\infty} \left|\frac{f(x)}{g(x)}\right| <\infty.
\end{equation}

Many of the ``common" clustering algorithms (e.g. K-means, BIRCH, STING, CURE, wavecluster, Particle Swarm Optimization, Quantum Dynamic Clustering, etc) 

have well established complexity classes.  Amongst the algorithms we are testing, the complexity classes are of the form
\[
a N^{b}\log^c(N)
\]

where $a$ may be a combination of factors such as dimension of data, branching threshold, minimum number of points, $\varepsilon$, etc.  With the size of the datasets we are scoring, none of these additional factors will be larger than $N^{\delta}$ where $\delta$ is some number less than one.  We in fact, test a branching factor where we have $a\approx N^{1/2}$. The logarithmic factor contributes very little to clock speed, so we score this as $N^{\gamma}$ where $\gamma$ is very small. Of course, when we consider the limit
\[
\forall c,\alpha>0,  \lim_{N\rightarrow \infty} \frac{\log^c(N)}{N^{\alpha}} = 0 
\]  

we can consider $\log^c(N)$ to be essentially constant.  However, in order to score our algorithms for speed, we need to set our reference point above our largest exponent.  The largest algorithm we have tested has the complexity
\[
O(aN^2\log^2(N)) \implies O(N^{2+1})
\]

Thus we have not seen an algorithm larger than $O(N^3)$ where $N$ is the size of the dataset.  We thus set our reference point at 4.  Our score for complexity now takes the form
\begin{equation}
\alpha\left(4 - \frac{\log(f(N))}{\log(N)}\right)^w
\end{equation}

Many of the algorithms we have tested clock-in around $O(N^2)$ thus we shall set $\alpha=50$.  Furthermore we asses complexity to be highly important, but not at the cost of all other features, thus we set $w=2$.  Giving us a score for complexity as

\begin{equation}
50\left(4- \frac{\log(f(N))}{\log(N)}\right)^2
\end{equation}

\subsection{Scalability and Dimensionality}

In some instances, one may wish to differentiate between row search and column search.  For a general algorithm, complexity takes these into account as a single quantity which is superceded by runtime complexity.  Thus for the current metric, we allow complexity to take this score into account. 

\section{Homogeneity}

By homogeneity we shall mean the homogeneity of a cluster after having determined by a clustering algorithm.  To accomplish this task we will work first on a single cluster.  Within this cluster we will determine how to score its individual homogeneity and then we will consider how to score the clusters in ensemble to gain our overall score.

We begin our task by considering the points in our cluster to be the vertices of a graph.  In this graph we will build the minimal spanning tree in which the weights are calculated by mutual reachability distance as in HDBSCAN\cite{HDBSCAN}.\\ 

Once we have built our minimal spanning tree we compute the mean $\mu$ and variance $\sigma^2$ of the weights of our minimal spanning tree.  Intuitively we guess that having a very small mean implies that our cluster is highly homogeneous.  The problem we run into is that we may have a dataset with highly varying densities in different regions.  Thus we shall also like to take variance into account.  Consider for a moment the following though experiments:

Consider three different clusters each with 100 points.  The first is the set of evenly spaced points on a circle of radius $r$, the second is the set of evenly spaced points on a cirle of radius $2r$, and the third a set of points on a cirle of radius $r$, but the spacings are random (Gaussian, with fixed variance).  Shown here

We see that the smaller circle with evenly spaced points is the most homogeneous.  It is now a question of how to rank the other 2.  Which is more homogeneous?  Are they the same?  

In order to answer this question we appeal to the Fisher information metric as in \cite{Stat-Man,Fisher}.  Given a set of random variables obeying some probability distribution we build a \emph{statistical} manifold from its parameter space.  Given that many datasets we will encounter have a large number of points $ >>1000$ we appeal to the central limit theorem and assume that our means and variances will obey a nearly normal distribution.  When we have a set $X$ with clusters $C_X$ and $|C_X|=N$ we construct the statistical manifold in the variables:
\[
(\theta_1,\theta_2,\dots,\theta_{2N-1},\theta_{2N}) = (\mu_1,\sigma^2_1,\dots,\mu_N,\sigma^2_N)
\]

with probability density function $f(x,\theta)$.

The canonical metric for a statistical manifold is given by
\begin{equation}
g_{\alpha\beta}(\theta) = \int_{\text{all space}} \frac{\partial \log (f)}{\partial \theta^{\alpha}}\frac{\partial \log (f)}{\partial \theta^{\beta}} f(x,\theta) dx
\end{equation}

In the case of Gaussian distributions the metric becomes
\begin{equation}
g = \begin{bmatrix}
1/\sigma^2_1 & &&&\\
 & 2/\sigma^2_1 &&&\\
 && \ddots && \\
 &&& 1/\sigma^2_N &\\
 &&&& 2/\sigma^2_N
\end{bmatrix}
\end{equation}

This gives us a length element of
\begin{equation}
d\ell^2 = \sum \frac{d\mu^2_i+ 2d\sigma^2_i}{\sigma^2_i}
\end{equation}

We recognize that this is the metric for the upper half space model of hyperbolic space, with some scaling factor on the variance.  In particular, for a single Gaussian distribution, we can see the scaling in distance becomes \cite{Fisher}
\begin{equation}
d_{\mathbb{F}}((\mu_1,\sigma_1),(\mu_2,\sigma_2)) = \sqrt{2}d_{\mathbb{H}}\left(\left(\frac{\mu_1}{\sqrt{2}},\sigma_1\right),\left(\frac{\mu_2}{\sqrt{2}},\sigma_2\right)\right)
\end{equation}

From our construction within clusters, we have the minimal spanning tree with weights determined by mutual reachability distance.
\[
\forall i \le |C_X|, T_i = (V_i,E_i,w_i)
\]

In this case we know that all distances and variances will be positive, and thus we wish to compute homogeneity as the fisher distance of $(\mu,\sigma)$ from the origin.  Unfortunately, every point is infinitely far from the origin by the construction of the metric. In order to get a meaningful measure, we shall scale up every variance one exactly one, and calculate distance of a cluster $(i)$ from $(0,1)$.  That is, we define the score $h_i$ by
\begin{equation}
h_i := d_{\mathbb{F}}\left((\mu_i,\sqrt{\sigma_i^2+1}),(0,1)\right)
\end{equation} 

Luckily much work has been done in this space and given $(\mu_i,\sigma_i)$ we have an explicit expression for $h_i$ as
\begin{equation}
h_i = \sqrt{2}\arcosh\left(1+\frac{\mu_i^2/2 + (\sqrt{\sigma_i^2+1}-1)^2}{2\sqrt{\sigma_i^2+1}}\right)
\end{equation}

\begin{rem}
This formula comes from the fact that the hyperbolic distance is given in general by
\[
d_{\mathbb{H}}((x_1,y_1),(x_2,y_2)) = \arcosh\left(1+\frac{(x_2-x_1)^2+(y_2-y_1)^2}{2y_1y_2}\right)
\]
\end{rem}

When $h_i$ is small, we have a highly homogeneous cluster, and when it is large, we have an inhomogeneous cluster. We define the dataset's homogeneity score $G$ by
\begin{equation}
G := 1+\sum_{j=1}^{N} h_j
\end{equation}

This gives us the bounds
\begin{equation}
G_{\min} = 1 < G \le 1+ N \max{h_j} = G_{\max}
\end{equation}

Recalling that $G_{\max}$ gives us many ($N$) highly inhomogeneous clusters we set this as our reference point $\beta$.  Now we wish to score things from zero to one-hundred, so we pick $\alpha = \frac{100}{G_{\max}}$.  Finally we give our weight factor as $w=1.1$.  We pick $w=1.1$ since it gives our homogeneity score a little more importance than simple linearity, but will give extraordinarily large values only in extreme cases when $G$ is very tiny and $G_{\max}$ is large.  Even so we're bounded above by
\[
\frac{100}{G_{\max}}\left(G_{\max}-G \right)^{1.1} < 100 G_{\max}^{0.1}
\]

\begin{equation}
\frac{100}{G_{\max}}\left(G_{\max}-G\right)^{1.1}
\end{equation}

\section{Intercluster Distance}

Intercluster distance is mildly important for most needs, but we can still glean some useful information from it.  If we have many large distances between clusters this means our algorithm separates clusters well.  For example, in a highly dense dataset $k$-means is likely to have very small separation between clusters, but Quantum Dynamic Clustering is likely to have a much large separation of clusters.

In order to define our intercluster distance, we consider the distance between two given clusters $C_1,C_2$.

\begin{equation}
d(C_1,C_2) := \min_{x\in C_1, y\in C_2} d(x,y)
\end{equation}

where $d:C_X\times C_X \rightarrow \R^+$ is an arbitrarily chosen metric suitable for one's purpose.  In our case we shall choose the simple Euclidean metric.
\[
d(x,y) = \sqrt{(x-y)^t(x-y)}, \: x,y\in \R^n
\]

Our overall distance $D$ for a chosen algorithm will be the sum of all intercluster distances.

\begin{equation}
D := \sum_{i<j} d(C_i,C_j)
\end{equation}

Note that we only need to consider $i<j$ since $d(C_i,C_i) = 0$ and as $d$ is a metric $d(C_i,C_j)=d(C_j,C_i)$.  For the sake of computational ease one can simply define and equivalent $D$
\[
D = \half\sum_{i=l}^{|C_X|}\sum_{j=1}^{|C_X|} d(C_i,C_j)
\]

The second distance score will be easier to code, although computationally twice as expensive.

We get the bounds:
\[
0\le D \le \binom{N}{2} d_{\max} = D_{\max}
\]

As there are $\binom{N}{2}$ pairs of clusters to consider.  If all clusters are the same distance apart, then we have exactly $D_{\max}=\binom{N}{2}d_{\max}$.

To bring this to a score within $[0,100]$ we consider our reference point to be $\beta=0$ since $k$-means will often produce a zero distance. In light of having a potential zero distance we shall scale our score by $\alpha = \frac{100}{1+D_{\max}}$.  Finally, we scale this by $w=1$.  This tells us that we will take the distance at ``face-value."  There is some importance to distance, but having a large intercluster distance does not overcome an algorithm's deficiencies in other scores.

\begin{equation}
\frac{100D}{1+D_{\max}}
\end{equation}

\section{Covolume}

In the current context, we shall take covolume to mean the amount of space which is not filled by our clusters.  This is consistent with the physical chemistry definition rather than the number theoretic definition.  Thus the scoring is rather straightforward, while the calculation may take a little effort.  

Consider a dataset $X$ which is a subset of $\R^n$.  We calculate the volume $V(X)$ as the volume of the convex hull of $X$.  Suppose now that $C_X$ is the set of clusters in $X$ which we enumerate as
\[
C_X = \{C_1,C_2,\dots,C_m \} 
\]  
The volume of each cluster $C_i$ is the volume of its convex hull.  Thus the covolume of $X$ is defined as
\begin{equation}
CoV(X) = V(X) - \sum_{i=1}^{m}V(C_i)
\end{equation}

We shall score this on as a percentage.  The higher the covolume the tighter the cluster bounds which gives us a score of type:
\[
\alpha \left(\frac{CoV(X)}{V(X)}\right)^w
\]

In this case we choose $\beta=0$ to be our reference point since we have with rare exception a covolume of zero from k-means clustering.  That is to say, it does not produce tightly bound clusters.  For our purposes we choose $\alpha=100$ since the theoretical maximum is near 1.  This will give us a maximum raw score of 100.  We recognize that having tightly bound clusters is not of utmost importance and thus we give this a smaller weight, $w=2$ which reduces the percentage. 

We then reveal our score for covolume as
\begin{equation}
100\left(\frac{CoV(X)}{V(X)}\right)^2
\end{equation}

\section{Shape}
The idea of ``shape" is complex enough that we only give an overview in this section, for a more complete treatment see \cite{NARG}\\

\par Once one has properly identified clusters, the metric of scoring how well an algorithm can classify an arbitrary shape arises.  This turns out to be amongst the most complicated metrics to score.  For example, many survey papers (cf give some references from sofya's data clustering book) say that an algorithm picks up an arbitrary shape or it does not.  We now ask the question, ``how arbitrary?" Can we quantify such a statement?  It is the belief of the authors that this too, can be computed in a consistent way.  

Our strategy is as follows:
\begin{enumerate}
\item Define clusters\\
\item Define the boundary points $(\partial C)$ of a cluster\\
\item Fit a ``smooth enough" $(C^2)$ curve or surface (or volume) to the boundary.\\
\item Compute the integral of curvature squared $(\int_S \kappa^2 d\sigma)$ over the surface.
\end{enumerate}

Defining the clusters is the work of whichever algorithm we are implementing.  After this we need to extract the boundary of the cluster.  Our initial thought was to compute the Minimal Spanning Tree of the cluster (in the mutual reachability metric) and define the boundary to be all vertices of degree one.  A moment of reflection, however will reveal many counterexamples, for example, a long, sparsely connected cluster with two endpoints near each other and all other vertices internal. (include a picture here).  A second though was to compute the minimal spanning tree of the entire graph and compute the boundary in the style of Cheeger (give reference)
\[
\partial S = \{(u,v)\in E(X) | u\in S, v\notin S \}
\]  
and picking out the vertices $u$ and $v$ as boundary points.  Again, this fails when considering a graph such as
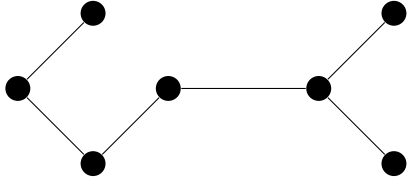
\[
\begin{tikzpicture}
\node[circle,fill](1) at (0,0){};
\node[circle,fill](2) at (1,1){};
\node[circle,fill](3) at (1,-1){};
\node[circle,fill](4) at (2,0){};
\node[circle,fill](5) at (4,0){};
\node[circle,fill](6) at (5,1){};
\node[circle,fill](7) at (5,-1){};
\foreach \from/\to in {1/2,1/3,3/4,5/6,5/7,4/5}
 \draw (\from) -- (\to);
\end{tikzpicture}
\]

In the example above we have only two boundary points.  This can't uniquely define a ``shape."  

In order to compute shape consistently we shall take the approach of polygonization of the boundary undertaken in \cite{LEC}.  This approach uses the DeLauney Triangulation of the cluster and extracts boundary points by oriented the edges in the triangulation.  This is similar to the approach of computing the shared nearest neighbor graph as in \cite{ESK}, but not with the intent of building a triangulation.  One may also wish to consider a barycentric division in higher dimensions.  Our algorithm will be to build a triangulation in 2 or 3 dimensions and extract the boundary in the style of \cite{LEC}.  

Once we have extracted the boundary, we need to build a surface which satisfies several criteria.  First, we would like the surface to encode some of the more exotic features of the data.  That is, we would like to build a curve or surface which is not simply a collection of line segments or planar regions hastily glued together to give a manifold with corners reminiscent of the early 1990s ``3D"  video games.  We will approach this task by the technique of B$\grave{e}$zier curves and surfaces \cite{Bez}.  

Once we have a surface with some curvature to it, and a reasonably good fit, we shall calculate the complexity of the shape of our cluster by computing the integral of its curvature squared.  This can be done efficiently for a curve \cite{NTGVB}.  For a surface we present the following calculation.

Let $S(u,v)$ be our surface parametrization

\[
S(u,v) = \sum_{ij} \mathbf{P}_{ij} N_{i4}(u)N_{j4}(v)
\]

The Gaussian ($K$) and mean ($H$) curvatures are defined as follow:

\begin{eqnarray}
K = \kappa_1 \kappa_2 & = & \frac{S_{uu}\cdot S_{vv} - S_{uv}^2}{\left(1+ S_u^2+S_v^2\right)^2}\\
H = \frac{\kappa_1+\kappa_2}{2} & = & \frac{(1+S_u^2)S_{vv} - 2S_u S_v S_{uv} +(1+S_v^2)S_{uu}}{\left(1+S_u^2+S_v^2\right)^{3/2}} 
\end{eqnarray} 

Then our curvature is given by the sum squared of principal curvatures 
\begin{equation}
\frac{\kappa_1^2+\kappa_2^2}{2} = 2H^2 - K
\end{equation}

Now we integrate

\begin{equation}
Shape(C) : =  \int_{u_0}^{u_1}\int_{v_0}^{v_1} (2H^2 - K) du dv
\end{equation}

Once we have scored each cluster $C$ we give the overall score by scaling as such

\begin{equation}
Shape := \alpha \left(\left(\sum_{C_i \in C_X} Shape(C_i)\right) - \beta\right)^w
\end{equation}

We wish our overall shape score to be as high as possible.  A high shape score signifies that an algorithm picks up an arbitrary shape with high precision.  For this reason, we shift our score back against the simplest possible shape to pick up, which is a circle (or sphere) of unit radius.  In two dimensions the curvature is constant $1/R$ and thus a constant positive unit.  Integrating over the circumference we get $2\pi$.  For a sphere we have two principal curvatures each of which is identically one thus we are integrating the constant function 2.  Integrating over the surface area we get $8\pi$.  Since most of our calculations will be over surfaces, we shall set $\beta = 8\pi$. This would say that an algorithm picking up a single cluster which is a uniform circular object scores as low as possible.  There is no practical limit to how high an individual cluster can score, so we scale by dividing by our maximum cluster score. $\alpha = \frac{1}{max_{i}(Shape(C_i))}$.  Finally, we regard this feature with utmost importance as to an algorithm's overall accuracy, and thus we shall weight it with $w=2$.  If necessary, a researcher should scale this back if the highest score is arbitrarily high.

\section{Our Scores}

\begin{table}[h!]

\begin{tabular}{l || c | c | c | c | c | c | c | c |}
Feature & $\alpha$ & $\beta$ & w & \multicolumn{2}{|c|}{Algorithm} & \multicolumn{3}{|c|}{Scores} \\
& & & & Best & Worst & Best & Worst & Average\\
\hline
& & & & & & & & \\
Stability & $\frac{100}{\sqrt{2}}$ & $\half$ & $\half$ & & & & &\\
& & & & & & & & \\
Noise sensitivity & $\frac{100}{\ln(dC_X)}$ & $\ln(dC_X)$ & 1.25 & & & & & \\
 & & & & & & & & \\
Complexity & 50 & 4 & 2 & & & & &\\
 & & & & & & & & \\
Homogeneity & $\frac{100}{G_{\max}}$ & $G_{\max}$ & $1.1$ & & & & &\\
 & & & & & & & & \\
Distance & $\frac{100}{1+D_{\max}}$ & $0$ & $1$ & & & & &\\
 & & & & & & & & \\
Covolume & 100 & 0 & 2 & & & & &\\
 & & & & & & & & \\
Shape & $\frac{1}{\max_i{Shape(C_i)}}$ & $8\pi |C_X|$ & 2 & & & & &\\
\end{tabular}

\caption{Table of empirical scores}
\label{table:1}
\end{table}


\newpage


\begin{thebibliography}{99}
\bibitem{Bench}\emph{Benchmarking Performance of Clustering Algorithms in Python:}
http://hdbscan.readthedocs.io/en/latest/performance\_and\_scalability.html


\bibitem{Scalability} \emph{A Fresh Graduate's Guide to Software Development Tools and Technique:Chapter 6, Scalability}, Khare, Huang, Doan, Kanwal, http://www.comp.nus.edu.sg/$\sim$seer/book/2e/Ch06.\%20Scalability.pdf

\bibitem{Gr-Cliu} \emph{A New graph-Theoretic Approach to Clustering and Segmentation},\\ 
M. Pavan, M. Pelillo,\\
Proceedings of the 2003 IEEE Computer Society on Computer Vision and Pattern Recognition\\
http://www.dais.unive.it/$\sim$pelillo/papers/cvpr03.pdf 

\bibitem{Stat-Man} \emph{Learning on Statistical Manifolds for Clustering and Visualization}\\
 K. Carter, R. Raich, A.O. Hero\\
http://tbayes.eecs.umich.edu/\_media/kmcarter/carter\_learnstatman.pdf

\bibitem{Fisher}\emph{Fisher Information Distance: A Geometrical Reading}\\
S. Costa, S. Santos, J. Strapason\\
https://arxiv.org/pdf/1210.2354.pdf

\bibitem{HDBSCAN} \emph{Hierarchical Density Estimates for Data Clustering, Visualization, and Outlier Detection} \\
R. Campello, D. Moulavi, A. Zimek, J. Sander\\ 
ACM Transactions on Knowledge Discovery from Data. 10 (1): 1–51. doi:10.1145/2733381.

\bibitem{ESK}\emph{Finding Clusters of Different Sizes, Shapes, and Densities in Noisy,
High Dimensional Data} 2003\\
 L. Ert\"{o}sz, M. Steinbach, V. Kumar\\
http://www-users.cs.umn.edu/$\sim$kumar/papers/SIAM\_snn.pdf 

\bibitem{LEC}\emph{Polygonization of Point Clusters through Cluster Boundary Extraction for Geographical Data Mining} I. Lee, V. Estivill-Castro\\
http://www.isprs.org/proceedings/XXXIV/part4/pdfpapers/098.pdf

\bibitem{NTGVB} \emph{Efficient Squared Curvature} C. Nieuwhuis, E. Toeppe, L. Gorelick, O. Vekslert, Yu. Boykov\\ http://www.csd.uwo.ca/$\sim$ygorelic/CVPR2014\_curvatureTR.pdf

\bibitem{Bez} \emph{B$\grave{e}$zier surfaces}\\ http://web.mit.edu/hyperbook/Patrikalakis-Maekawa-Cho/node14.html

\bibitem{NARG} \emph{The Shape Metric for Clustering Algorithms}\\ C. Alexander, S. Akhmametyeva, 2017

\end{thebibliography}
\end{document}